\documentclass[12pt]{amsart}
\usepackage{amsmath}
\usepackage{amssymb}
\usepackage[latin1]{inputenc}
\usepackage[T1]{fontenc}
\newtheorem{lemma}{Lemma}
\newtheorem{theorem}{Theorem}
\newtheorem{proposition}{Proposition}
\newcommand{\iso}{{\rm Iso}}
\newcommand{\tr}{{\rm Tr }}
\newcommand{\wG}{\tilde{\Gamma}}
\newcommand{\tG}{\Gamma}
\newcommand{\hoc}[1]{\gamma(#1)}
\newcommand{\ho}[1]{\alpha(#1)}
\newcommand{\hol}{\alpha}
\newcommand{\holc}{\gamma}
\newcommand{\hog}[1]{\beta(#1)}
\newcommand{\holg}{\beta}
\newcommand{\aut}{{\rm Aut }}
\newcommand{\Cc}{\h C_c}
\newcommand{\intL}[2]{L(#1,#2)}
\newcommand{\intd}[2]{D(#1,#2)}
\newcommand{\bra}{\langle}
\newcommand{\ket}{\rangle}
\newcommand{\vp}{\varphi}

\newcommand{\C}{\mathbb{C}}
\newcommand{\N}{\mathbb{N}}
\newcommand{\Q}{\mathbb{Q}}
\newcommand{\R}{\mathbb{R}}
\newcommand{\be}{\begin{equation}}
\newcommand{\eeq}{\end{equation}}
\newcommand{\bet}{\begin{equation*}}
\newcommand{\eeqt}{\end{equation*}}
\newcommand{\bea}{\begin{eqnarray}}
\newcommand{\eeqa}{\end{eqnarray}}
\newcommand{\beat}{\begin{eqnarray*}}
\newcommand{\eeqat}{\end{eqnarray*}}
\newcommand{\goesto}{\longrightarrow}
\newcommand{\h}[1]{\mathcal{#1}}
\newcommand{\hil}{\mathcal{H}}
\newcommand{\uop}{\mathcal{O}(\hil)}

\newcommand{\hT}{\mathcal{T}}

\newcommand{\hD}{\mathcal{D}}
\newcommand{\hA}{\mathcal{A}}
\newcommand{\hB}{\mathcal{B}}

\newcommand{\hF}{\mathcal{F}}

\begin{document}
\title{Normal covariant quantization maps}
\author{J. Kiukas}
\address{Jukka Kiukas,
Department of Physics, University of Turku,
FIN-20014 Turku, Finland}
\email{jukka.kiukas@utu.fi}
\author{P. Lahti}
\address{Pekka Lahti,
Department of Physics, University of Turku,
FIN-20014 Turku, Finland}
\email{pekka.lahti@utu.fi}
\author{K. Ylinen}
\address{Kari Ylinen,
Department of Mathematics, University of Turku,
FIN-20014 Turku, Finland}
\email{kari.ylinen@utu.fi}
\begin{abstract}
We consider questions related to quantizing complex valued functions defined on a locally compact topological
group. In the case of bounded functions, we generalize R. Werner's approach to prove the characterization
of the associated normal covariant quantization maps.
\end{abstract}
\maketitle

\section{Introduction}

Quantization is a procedure which associates a quantum mechanical observable to a given
classical dynamical variable, the latter being represented by a complex valued Borel function on the phase space
$X$ of a classical system. The phase space $X$ can be taken to be $\R^{2n}$ or, more generally, $G / H$, where $G$
is a locally compact second countable topological group and $H$ a closed subgroup. We consider here only the case
where $X=G$. Quantization can be realized e.g.
by integrating the classical variable $f$ with respect to a suitable (positive normalized)
operator measure $E:\hB(G)\to L(\hil)$, where $\hB(G)$ is the Borel $\sigma$-algebra of subsets of $G$ and
$L(\hil)$ the set of bounded operators acting on the Hilbert space $\hil$ of the quantum system.
The resulting operator integral $L(f,E)=\int f dE$ is a (possibly unbounded) linear operator, which is symmetric
if $f$ is real valued. (See Section \ref{sec6} for our definition of the domain of the operator integral.) In many cases,
the operator $L(f,E)$ is essentially selfadjoint, so that it is eligible to represent a quantum observable.
The map $f\mapsto L(f,E)$ is linear (in the sense made precise in Section \ref{sec6}). If $f$ is bounded, then $L(f,E)\in L(\hil)$.
The operator integral has a convergence property, which could be called ''quasicontinuity'' (see e.g. \cite[p. 22]{Berberian}):
If $(f_n)$ is an increasing sequence of positive Borel functions converging pointwise to a Borel function $f$,
and $\vp\in \hil$ is a vector belonging to the domains of $L(f,E)$ and each $L(f_n,E)$, then the sequence
$(\bra \psi|L(f_n,E)\vp\ket)$ of numbers converges for each $\psi\in\hil$ to $\bra \psi|L(f, E)\vp\ket$.

As noted above, quantization might be any mapping $\Gamma$ from the set of Borel functions
to the set of linear operators on $\hil$. It is therefore natural to ask which of them can be represented by
operator integrals with respect to some positive
operator measures. Essential requirements for $\Gamma$ are linearity, positivity, the property that bounded
functions are mapped to $L(\hil)$, and quasicontinuity, as
they assure that the association $B\mapsto \Gamma(\chi_B)$ defines a positive operator measure $E^{\Gamma}$.
Obviously, this does not guarantee that the quantization map $\Gamma$ would coincide
with the map given by the operator integral with respect to $E^{\Gamma}$; in particular, nothing has been said about
the domains of the operators $\Gamma(f)$. In the case of a bounded function $f$, however, the domain of the operator
integral $L(f, E^{\Gamma})$ is all of $\hil$, and it follows easily
that $\Gamma(f) = L(f, E^{\Gamma})$. Thus, if we have a positive and quasicontinuous linear
quantization map $\Gamma$, which maps bounded functions to $L(\hil)$, then (at least) the restriction of $\Gamma$
to the set of bounded functions can be represented as the operator integral $L(\cdot, E^{\Gamma})$. 

Since the phase space $G$ has a left Haar measure $\lambda$, it is convenient to consider the functions in
$L^\infty(G,\lambda)$ (i.e. $\lambda$-equivalence classes of $\lambda$-essentially bounded $\lambda$-measurable
complex functions) instead of bounded Borel functions. Assume that the original quantization map $\Gamma$ (defined on
all complex Borel functions) is linear, positive, has the quasicontinuity property, and maps bounded functions to $L(\hil)$.
In addition, we can require that each complex measure $B\mapsto E^{\Gamma}_{\psi,\vp}(B) = \bra \psi |E^{\Gamma}(B)\vp\ket$
is $\lambda$-continuous. This ensures that $\Gamma(f)$ does not
depend on the (Borel) representative of $f\in L^\infty(G,\lambda)$, so we get a well-defined positive linear quantization map
$\tilde{\Gamma}:L^\infty(G,\lambda)\to L(\hil)$ which coincides with the map obtained from $L(\cdot, E^{\Gamma})$ in the
similar way.

When we restrict our attention to the positive linear quantization maps $\Gamma$ defined on $L^\infty(G,\lambda)$ with values in
$L(\hil)$, the condition of quasicontinuity is not appropriate, as it involves pointwise convergence. Instead, we require the
somewhat similar condition of normality,
i.e. weak-* continuity associated with the dualities $L^1(G,\lambda)^*=L^\infty(G,\lambda)$ and $\hT(\hil)^*=L(\hil)$,
where $L^1(G, \lambda)$ is the set of $\lambda$-equivalence classes of $\lambda$-integrable
complex functions and $\hT(\hil)$ is the set of all trace class operators on $\hil$.
Thus, if the $\lambda$-continuity of the complex measures
$E^{\Gamma}_{\psi,\vp}$ is assumed, we have $\Gamma = L(\cdot, E^{\Gamma})$. Conversely, if
a positive operator measure $E$ is given, for which each $E_{\psi,\vp}$ is $\lambda$-continuous, then the map
$L^\infty(G,\lambda)\ni f\mapsto L(f,E)\in L(\hil)$ is linear, positive, and normal (see Section \ref{sec5}).

An important property of a quantization map $\Gamma: L^\infty(G,\lambda)\to L(\hil)$
(or the corresponding operator measure) is covariance (see Section \ref{sec2}),
which connects it to the structure of the phase space. Covariance also conveniently implies the $\lambda$-continuity
of the complex measures $E^{\Gamma}_{\psi, \vp}$ (see section \ref{sec5}).
Covariant positive phase space operator measures have proved highly useful also in various other applications of quantum mechanics, including
for instance the fundamental questions on joint measurements of position and momentum observables and the problem
of quantum state estimation (quantum tomography). Consequently, the structure of such operator measures has been studied extensively:
the canonical examples of the covariant phase space observables are constructed e.g. in \cite{Davies}, whereas a complete
group theoretical characterization is given in \cite{Cassinelli}.

The characterization of \cite{Cassinelli} is based on a generalization of Mackey's imprimitivity theorem \cite{Cattaneo}. However,
in the concrete case where the phase space is $\R^{2n}$, there is another, more direct (and completely different) approach, outlined by Holevo
\cite{Holevo}, and further elaborated by Werner \cite{Werner}. In fact, in \cite{Werner}, Werner
characterizes all the positive normal phase space translation covariant maps $\Gamma: L^\infty(\R^{2n})\to L(\hil)$.
The essential part of both Holevo's and Werner's proofs relies on the fact that the Banach space of trace class operators on a separable Hilbert space
has the Radon-Nikod\'ym property.

In this paper we generalize Werner's approach to the case where the phase space is a locally compact unimodular
topological group, paying due attention to the details arising in this context. In addition, we consider briefly
the question of quantization of unbounded functions.

\section{Preliminaries}\label{sec2}

If $\hil$ is a Hilbert space, we let $L(\hil)$ and $\hT(\hil)$ denote the sets of bounded operators and trace class
operators on $\hil$, respectively. 

Let $\mu_L$ denote the Lebesgue measure of $\R^{2n}$.
Denote the Weyl operators on the Hilbert space $L^2(\R^n)$ by $W(x)$, $x = (q,p)\in \R^{2n}$, so that $W(q,p)$ acts according to
\bet
(W(q,p)\psi)(t) = e^{i\frac 12 q\cdot p}e^{ipt}\psi(t+q).
\eeqt
They satisfy the relation
\be\label{proj}
W(x)W(y)= e^{i\frac 12 \{x, y\}}W(x+y),
\eeq
where  $\{(q,p), (q',p')\}= q\cdot p'-p\cdot q'$ for all $(q,p), (q',p')\in \R^{2n}$.

For each $x\in \R^{2n}$, define $\hoc{x}:\hT(L^2(\R^n))\to \hT(L^2(\R^n))$ by $\hoc{x}(T) = W(x)TW(-x)$. Then the map
$x\mapsto \holc(x)$ has the following well-known properties. The proof is included for the reader's convenience.
\begin{lemma}\label{easylemma}
\begin{itemize}
\item[(a)] $\hoc{x+y} = \hoc{x} \circ \hoc{y}$ for all $x, y\in \R^{2n}$.
\item[(b)] $\hoc{x}^*(A) = W(-x)AW(x)$ for all $A\in L(L^2(\R^n))$ and $x\in \R^{2n}$.
\item[(c)] $\hoc{x}$ is a positive trace-norm isometry for all $x\in \R^{2n}$.
\item[(d)] For each $A\in L(L^2(\R^n))$ and $S\in \hT(L^2(\R^n))$, the function $x\mapsto \tr[A\hoc{x} (S)]$ is continuous.
\item[(e)] $\int \tr[P_1 \hoc{x}(P_2)] d\mu_L(x)=(2\pi)^n$ for all one-dimensional projections $P_1$ and $P_2$ on $L^2(\R^n)$.
\end{itemize}
\end{lemma}
\noindent {\bf Proof. } (a) is a direct consequence of the relation (\ref{proj}), and (b) follows from a basic property of the trace.
If $U$ is a unitary operator, $|USU^*| = U|S|U^*$ for each $S\in L(L^2(\R^n))$.
Therefore, since $W(x)$ is unitary and $W(x)^*=W(-x)$, $\| \hoc{x} (S)\|_{\rm tr} = \tr [|W(x)SW(-x)|] = \|S\|_{\rm tr}$ for each
$S\in \hT(L^2(\R^n))$. This proves (c), as it is clear that $\hoc{x}$ is positive.
To prove (d), take $A\in L(L^2(\R^n))$ and $S\in \hT(L^2(\R^n))$. Let $x\in \R^{2n}$, and $(x_n)$ be a sequence converging to $x$.
Since $x\mapsto W(x)$ is strongly continuous, $\hoc{x_n}^*(A) = W(-x_n)AW(x_n)\goesto W(-x)AW(x)=\hoc{x}^*(A)$ weakly.
Since all $W(x)$ are unitary, the sequence $(\hoc{x_n}^*(A))$ is norm bounded, from which it follows that it converges
to $\hoc{x}^*(A)$ also ultraweakly. Thus we get
\bet
\tr[A\hoc{x_n} (S)] = \tr[\hoc{x_n}^*(A)S]\goesto \tr[\hoc{x}^*(A)S] = \tr [A\hoc{x} (S)],
\eeqt
which proves (d). The proof of (e) goes as follows. Assume that $P_1=|\psi\ket\bra\psi|$ and $P_2=|\vp\ket\bra\vp|$,
where $\psi,\vp\in\hil$ are unit vectors. Define the function $\phi_q$ for each $q\in\R^n$ by $\phi_q(t)=\psi(t)\vp(t+q)$. Then
\bet
1=\|\psi\|^2\|\vp\|^2 = \int \left(\int |\psi(t)|^2|\vp(q)|^2 dq\right)dt = \int \left(\int |\psi(t)|^2|\vp(t+q)|^2 dt\right)dq
\eeqt
by the Fubini-Tonelli theorem, so that $\phi_q\in\hil$ for almost all $q$. By the unitarity of the inverse
Fourier-Plancherel operator $F$, we have now
\bet
1= \int \int |(F\phi_q)(p)|^2 dpdq.
\eeqt
But since $\psi$ and $\vp(\cdot+q)$ are in $L^2(\R^n)$, $\phi_q$ is also integrable, so
\bet
(F\phi_q)(p)=\frac {1}{\sqrt{(2\pi)^n}}\int e^{ip\cdot t}\phi_q(t)dt
= \frac {1}{\sqrt{(2\pi)^n}} e^{-i\frac 12p\cdot q}\bra \psi|W(q,p)\vp\ket,
\eeqt
from which it follows that
\bet
(2\pi)^n=\int |\bra \psi|W(x)\vp\ket|^2 d\mu_L(x)= \int \tr[P_1 \hoc{x}(P_2)] d\mu_L(x).
\eeqt
$\Box$

\

Now we proceed to a more abstract case.

If $(\Omega, \hA,\nu)$ is a $\sigma$-finite (positive) measure space, we let
$L^1(\Omega, \nu)$ denote the Banach space of (equivalence classes of) complex valued,
$\nu$-integrable functions, and $L^\infty(\Omega, \nu)$ the Banach space of (equivalence classes of) complex valued,
$\nu$-measurable, $\nu$-essentially bounded functions.

A function $g$ defined on $\Omega$ and having values in
some Banach space is said to be $\nu$-measurable, if for each $B\in \hA$ of finite measure there is a sequence of
$\nu$-simple functions converging to $\chi_Bg$ in $\nu$-measure (or, equivalently, there is a sequence
of $\nu$-simple functions which converges $\nu$-almost everywhere to $\chi_Bg$) \cite[pp. 106, 150]{Dunford}.
In the case where the value space of $g$ is separable (in particular, if $g$ is scalar-valued), $\nu$-measurability is
equivalent to the measurability with respect to the Lebesgue extension of the $\sigma$-algebra $\hA$ with respect
to $\nu$ \cite[p. 148]{Dunford}.
If $X$ is a Banach space, $\iso(X)$ denotes the group of linear homeomorphisms from $X$ onto $X$.

Let $\hil$ be a separable Hilbert space. Let $\aut (\hT(\hil))$ denote the subgroup of $\iso(\hT(\hil))$
consisting of the positive maps which preserve the trace norm. The set $\aut(\hT(\hil))$ is equipped with the weak topology
given by the set of functionals $u\mapsto \tr[A u(T)]$, where $A\in L(\hil)$, $T\in \hT(\hil)$.
For $u\in \aut(\hT(\hil))$, the adjoint map $u^*:L(\hil)\to L(\hil)$ restricted to $\hT(\hil)$ is equal to $u^{-1}$.
It follows from the Wigner theorem that for each $u\in \aut(\hT(\hil))$ there is an either
unitary or antiunitary operator $U$, such that
$u(T) = UTU^*$ for all $T\in\hT(\hil)$.

Let $G$ be a locally compact unimodular second countable (Hausdorff) topological group, with Haar measure $\lambda$,
such that there is a continuous group homomorphism $\holg:G\to \aut(\hT(\hil))$ and a constant $d>0$, satisfying
\be\label{prop1}
\int \tr[P_1\hog{g}(P_2)] d\lambda(g) = d
\eeq
for all one-dimensional projections $P_1$ and $P_2$ on $\hil$. The system $(G, \holg, d)$
will remain fixed throughout the paper.

\noindent {\bf Remark. }
\begin{itemize}
\item[(a)] It follows from Lemma \ref{easylemma} that the additive group $\R^{2n}$, with the homomorphism
$\holc$ and the constant $(2\pi)^n$ constitute an example of the abstract system $(G, \holg, d)$.
\item[(b)]
The fact that each $\hog{g}$ has the form $\hog{g}(T) = U(g)TU(g)^*$ for some unitary or antiunitary operator $U(g)$ implies that,
in the case where $G$ is connected, the map $g\mapsto U(g)$ is a projective unitary representation of $G$ 
which satisfies the square integrability condition
\bet
\int |\bra \psi|U(g)\vp\ket|^2 d\lambda(g) = d
\eeqt
for all unit vectors $\psi,\vp\in\hil$. The theory of such representations and the associated covariant operator
measures is well developed, see e.g. \cite{Ali}. It can be noted that in the case of a nonunimodular group, the
square integrability condition is no longer of the above form for some fixed $d$ \cite{Duflo}.

In this paper, however, we do not need the explicit structure of the map $\holg$ given by the projective representation $g\mapsto U(g)$.
Thus we will use only the abstract definition, with the map $\holc$ associated with the group $\R^{2n}$ as a
concrete example.
\end{itemize}

\

If $S$ is a positive trace class operator and $A$ a bounded positive operator, the function
$G\ni g\mapsto \tr [A\hog{g}(S)]$ is positive.
Concerning the integrability of such a function, the following lemma
holds (with the understanding that $\infty\cdot 0 = 0$):
\begin{lemma}\label{trlemma}
Let $S\in\hT(\hil)$ and $A\in L(\hil)$ be positive operators. Then
\bet
d^{-1}\int \tr [A\hog{g} (S)] d\lambda(g) = \tr [A] \ \tr [S].
\eeqt
In particular, if $S\neq O$, the function $g\mapsto \tr [A\hog{g} (S)]$ is integrable if and only if $A\in \hT (\hil)$. 
\end{lemma}
\noindent {\bf Proof.} The proof consists of several stages.
\begin{itemize}
\item[1)] Assume that $A$ and $S$ are one-dimensional projections. Since now $\tr[A] \ \tr[S]=1$, it follows
directly from the relation (\ref{prop1}) that $d^{-1}\int \tr [A\hog{g} (S)] d\lambda(g) = \tr [A] \ \tr [S]$.
\item[2)] Assume that $S$ is a positive nonzero trace class operator and $A$ a one-dimensional projection.
Then $S = \sum_{i=1}^\infty w_i |\vp_i\ket\bra \vp_i|$,
where $(\vp_i)$ is an orthonormal sequence in $\hil$, the series converging in the trace norm, and
$w_i\geq 0$, $\sum_i w_i = \tr [S]$.
Since the map $T\mapsto \tr [A\hog{g} (T)]$ is linear and trace-norm continuous, we have
$\tr [A \hog{g} (S)] = \sum_i w_i \tr [A\hog{g}(|\vp_i\ket\bra \vp_i|)]$ for all $g$. Now
the result 1) and the monotone convergence theorem give
\beat
\tr [A] \ \tr [S] &=& \sum_i w_i \tr[A]\tr [|\vp_i\ket\bra \vp_i|] = \sum_i w_i d^{-1}\int \tr [A\hog{g}(|\vp_i\ket\bra \vp_i|)] d\lambda(g)\\
&=& d^{-1}\int \tr [A\hog{g} (S)] d\lambda(g).
\eeqat
\item[3)] Let $S$ be as before, and $A$ a bounded positive operator such that the set $\sigma_p(A)$
of eigenvalues of $A$ equals either the spectrum $\sigma(A)$ or the set $\sigma(A)\setminus \{0\}$.
(In particular, all the positive compact operators are like this.) Now $E^A(\sigma_p(A)) = I$, where $E^A$ is the spectral
measure of $A$.
It follows that the eigenvectors of $A$ constitute an orthonormal basis of $\hil$. Since $\hil$ is separable,
the set $\sigma_p(A)$ is at most countable. Let $\sigma_p(A) = \{a_1, a_2,\ldots\}$, and
$(\vp_{nk})$ be an orthonormal basis of $\hil$, such that for each $n$, the vectors $\vp_{nk}$ span the eigenspace of
$A$ associated with the eigenvalue $a_n$. Now $\tr [A] = \sum_{nk} \bra \vp_{nk} |A\vp_{nk}\ket = \sum_n a_n d_n$,
where $d_n\leq\infty$ is the degree of the eigenvalue $a_n$. Moreover,  
\bet
\tr [A\hog{g} (S)] = \sum_{nk} a_n \bra \vp_{nk}|\hog{g}(S)\vp_{nk}\ket
= \sum_{nk} a_n \tr [|\vp_{nk}\ket\bra \vp_{nk}|\hog{g}(S)].
\eeqt
It now follows from 2) and the monotone convergence theorem that
\bet
d^{-1}\int \tr [A\hog{g} (S)] d\lambda(g) =\sum_n a_n d_n \tr[S]= \tr [A] \ \tr [S].
\eeqt
In particular, if $A$ has an eigenspace of infinite dimension corresponding to a nonzero eigenvalue, then
$\int \tr [A\hog{g} (S)] d\lambda (g)= \infty$.
\item[4)] Let again $S$ be a positive nonzero trace class operator. Assume that $A$ is a positive bounded
operator, such that the set of
eigenvalues of $A$ equals neither the whole spectrum $\sigma(A)$ nor the set $\sigma(A)\setminus \{0\}$. Then
$\sigma(A)$ contains a point $a_0>0$, which is not an eigenvalue of $A$. Now $E^A(\{a_0\}) = O$.
It follows that $E^A(I_\epsilon)$, where $I_\epsilon = (a_0-\epsilon, a_0+\epsilon)$, is infinite-dimensional for all
$\epsilon >0$.
Let $t = \frac {a_0}{2}$. Then $t\chi_{I_t}(x)\leq x$ for all
$x\geq 0$, so that $tE^A(I_t) = \int t\chi_{I_t}(x) dE^A(x) \leq \int x dE^A(x) = A$.
Since $E^A(I_t)$ is an infinite dimensional projection, $\infty = t\tr [E^A(I_t)]\leq \tr [A]$, and hence also $\tr [A] = \infty$.
In addition, since $\hog{g}(S)$ is positive, $t\tr [E^A(I_t)\hog{g}(S)]\leq \tr [A\hog{g} (S)]$.
Since the projection $E^A(I_t)$ is infinite-dimensional, 3) implies that the function $g\mapsto t\tr[E^A(I_t)\hog{g}(S)]$
is not integrable. Thus $d^{-1}\int \tr [A\hog{g} (S)] d\lambda(g) = \infty = \tr [A] \tr [S]$.
\end{itemize}
The lemma is proved. $\Box$

\

In the following definition, note that the class of the function $f(g\cdot)\in L^\infty(G,\lambda)$ is independent of
the representative of $f$.

\

\noindent {\bf Definition.} A linear map $\Gamma:L^{\infty}(G, \lambda)\to L(\hil)$ is said to be \emph{$\holg$-covariant}, if
$\hog{g}^*(\Gamma(f)) =\Gamma(f(g\cdot))$ for all $f\in L^\infty(G, \lambda)$, $g\in G$. 

\

The main result in this paper, Theorem \ref{Theorem2}, has the following rather straightforward and, at least in special
cases, well-known converse.

\begin{theorem}\label{Theorem1}
Let $T$ be a positive operator of trace one. Then for each $f\in L^\infty(G, \lambda)$, the integral
\be\label{intT}
d^{-1}\int f(g) \hog{g} (T)d\lambda(g)
\eeq
exists as an operator $\Gamma(f)\in L(\hil)$ in the ultraweak sense. In addition, $\Gamma(g\mapsto 1) =I$, and the map
$f\mapsto \Gamma(f)$ is linear, positive, normal, and $\holg$-covariant.
\end{theorem}
\noindent {\bf Proof. } By Lemma \ref{trlemma}, the function $g\mapsto \tr [S\hog{g} (T)]$ is in $L^1(G, \lambda)$ for each
trace class operator $S$ (the operator $S$ can be written as a linear combination of four positive trace-class operators).
Thus for each $f\in L^\infty(G, \lambda)$ we can define the (clearly linear) functional $\Phi_f: \hT(\hil)\to \C$ by
\bet
\Phi_f(S) = d^{-1}\int f(g) \tr [S\hog{g} (T)] d\lambda(g).
\eeqt
Let now $f\in L^\infty(G, \lambda)$ be real valued. If $S\in \hT(\hil)$ is positive,
we have by Lemma \ref{trlemma}
\bet
|\Phi_f(S)|\leq d^{-1}M_f \int \tr [S\hog{g} (T)] d\lambda(g) = M_f \| S\|_{\rm tr},
\eeqt
where $M_f <\infty$ is such that $f(g)\leq M_f$ for almost all $g$. If $S\in \hT(\hil)$ is selfadjoint,
it can be written in the form $S = S^+-S^-$, where $S^{\pm}\in\hT(\hil)$ are positive and $|S| = S^++S^-$. Now
\bet
|\Phi_f(S)| \leq |\Phi_f(S^+)|+|\Phi_f(S^-)|\leq M_f (\| S^+\|_{\rm tr}+\| S^-\|_{\rm tr}) =M_f \|S\|_{\rm tr},
\eeqt
so that for real valued $f$, the map $\Phi_f$ restricted to the set of selfadjoint trace class operators is a real valued
trace-norm continuous linear functional. Hence, there is a selfadjoint operator $\Gamma(f)\in L(\hil)$, such that
$\Phi_f(S) = \tr [S \Gamma(f)]$ for all selfadjoint $S\in \hT(\hil)$. For an arbitrary $S\in\hT(\hil)$, we have $S = S_1+iS_2$, 
where each $S_i\in \hT(\hil)$ are selfadjoint, and so
\bet
\Phi_f(S)= \Phi_f(S_1)+i\Phi_f(S_2) = \tr [S_1\Gamma(f)]+i\tr [S_2\Gamma(f)]= \tr [S\Gamma(f)].
\eeqt
Let now $f\in L^\infty(G, \lambda)$ be complex valued: $f=f_1+if_2$, where $f_1$ and $f_2$ are real. Then clearly
$\Phi_f(S) = \Phi_{f_1}(S)+i\Phi_{f_2}(S) = \tr [S (\Gamma(f_1)+i\Gamma(f_2))]$ for all $S\in\hT(\hil)$. Define
$\Gamma(f) := \Gamma(f_1)+i\Gamma(f_2)\in L(\hil)$. Now
\bet
\Phi_f(S) = \tr[S\Gamma(f)]
\eeqt
for all $S\in\hT(\hil)$ and $f\in L^\infty(G, \lambda)$, implying the existence of the integral (\ref{intT}) in the ultraweak sense as
the operator $\Gamma(f)\in L(\hil)$.

The statement $\Gamma(g\mapsto 1) = I$ follows from Lemma \ref{trlemma}. 

Clearly $\Gamma: L^\infty(G, \lambda)\to L(\hil)$ is linear. If $f\geq 0$ and $\vp\in\hil$,
\bet
\bra \vp|\Gamma(f)\vp\ket = \tr [|\vp\ket\bra \vp|\Gamma(f)] = \Phi_f(|\vp\ket\bra \vp|)
= d^{-1}\int f(g)\bra\vp |\hog{g} (T)\vp\ket d\lambda(g) \geq 0,
\eeqt
which proves the positivity of $\Gamma$. Since $\Phi_f(S) = \tr[S\Gamma(f)]$ for all $S\in\hT(\hil)$ and $f\in L^\infty(G, \lambda)$,
$\Gamma$ is the dual of the map $\hT(\hil)\ni S\mapsto d^{-1}\tr [S\hog{\cdot} (T)]\in L^1(G, \lambda)$, and hence normal.

Covariance is seen from the calculation 
\beat
\tr [S\Gamma(f(g\cdot))] &=& d^{-1}\int f(gg') \tr[S\hog{g'}(T)] d\lambda(g') = d^{-1}\int f(g') \tr [S\hog{g^{-1}g'}(T)] d\lambda(g') \\
&=& d^{-1}\int f(g') \tr[\hog{g} (S)\hog{g'} (T)] d\lambda(g') = \tr [\hog{g} (S) \Gamma(f)] = \tr [S \hog{g}^* (\Gamma(f))],
\eeqat
where $g\in G$ and $f\in L^\infty(G,\lambda)$ are arbitrary, and the left invariance of $\lambda$, along with the properties
of the map $\holg$, are used. $\Box$

\section{General covariant maps}\label{sec3}

In this section we formulate the essential part of the characterization in yet a slightly more general context.
Let $X$ a Banach space having the Radon-Nikod\'ym property, i.e., if $(\Omega, \hA, \nu)$ is a finite (positive)
measure space and $\mu:\hA\to X$ a $\nu$-continuous vector measure with bounded variation,
there is a $\nu$-(Bochner-)integrable function
$g_\mu:\Omega\to X$, such that $\mu(B) = \int_B g_{\mu} d\nu$ for all $B\in \hA$ (see \cite[p. 61]{Diestel}).
The function $g_\mu$ is $\nu$-essentially unique \cite[p. 47, Corollary 5]{Diestel}.

The statement of the following Lemma is called the Riesz Representation Theorem. It is proved in
\cite[pp. 62-63]{Diestel}, in the case where $\nu$ is a finite measure. The Lemma here is an obvious generalization of
that result to the $\sigma$-finite case. As it constitutes the very starting point of the proof of the main result of the
paper, we give its proof here.

\begin{lemma} \label{oplemmaA}
Let $(\Omega, \hA, \nu)$ be a $\sigma$-finite measure space, $X$ a Banach space having the Radon-Nikod\'ym
property, and $\tG:L^1(\Omega, \nu)\to X$ a continuous linear map. Then there is a $\nu$-essentially
unique $\nu$-measurable function $v:\Omega\to X$, such that $\sup_{x\in \Omega}\|v(x)\| = \|\tG\|$, and
\bet
\tG(f) = \int f v d\nu
\eeqt
for all $f\in L^1(\Omega, \nu)$.
\end{lemma}
\noindent {\bf Proof. }
Choose a disjoint sequence $(K_n)$ of sets in $\hA$, such that $\Omega=\bigcup_n K_n$, and $\nu(K_n)<\infty$.
Denote by $\nu_n$ the restriction of $\nu$ to the $\sigma$-algebra $\hA(K_n)=\{B\cap K_n | B\in \hA\}$.
Define the set function $\mu_n:\hA(K_n)\to X$ by $\mu_n(B) = \tG(\chi_B)$. Now
$\|\mu_n(B)\| \leq \|\tG\|\|\chi_B\|_1=\|\tG\|\nu_n(B)$ for all $B\in \hA(K_n)$.
It follows that $\mu_n$ is a $\nu_n$-continuous vector measure of bounded variation, with the variation satisfying
$|\mu_n|(B)\leq \|\tG\| \nu(B)$ for all $B\in \hA(K_n)$.
Since $X$ has the Radon-Nikod\'ym property and $\nu_n$ is a finite measure,
there is a $\nu_n$-integrable function
$v_n:K_n\to X$, such that $\mu_n(B) =\int_B v_n d\nu_n$ for all $B\in \hA (K_n)$.
For each $f\in L^1(K_n, \nu_n)$, let $\tilde{f}$ be the function $\Omega\to \C$ which coincides with $f$ in $K_n$ and is
zero elsewhere. Since the map $L^1(K_n, \nu_n)\ni f\mapsto \tG(\tilde{f})\in X$
is linear and continuous, it follows from \cite[Lemma 4, p. 62]{Diestel} that $\|v_n(x)\| \leq \|\tG\|$ for $\nu_n$-almost all $x\in K_n$,
and $\tG(\tilde{f}) = \int f v_n d\nu_n$ for each $f\in L^1(K_n, \nu_n)$. By \cite[Corollary 5, p. 47]{Diestel}, $v_n$ is
$\nu_n$-essentially unique, and $v_n$ can be redefined to be zero in the null set in which originally $\|v_n(x)\|> \|\tG\|$.
Now we have $\sup_{x\in K_n} \|v_n(x)\|\leq \|\Gamma\|$. 

We denote by $v_n$ also the function $\Omega\to X$ which coincides with $v_n$ in $K_n$ and is zero
elsewhere. Since the sets $K_n$ are disjoint, we can define $v=\sum_n v_n$, where the sum converges pointwise. 
Since $v$ is a pointwise limit of $\nu$-measurable functions, it is itself $\nu$-measurable.
Denote $M= \sup_{x\in \Omega} \|v(x)\|\leq \|\Gamma\|$.

Let $f\in L^1(\Omega, \nu)$. Now the sequence $(f_k)$, where $f_k = \chi_{\cup_{n=1}^k K_n}f$ converges pointwise, and hence
(by the dominated convergence theorem) in $L^1(\Omega, \nu)$ to $f$. By continuity, $(\tG(f_k))$
converges to $\tG(f)$ in $X$.
On the other hand, since $\|f_k(x) v(x)\| \leq |f(x)|\|\tG\|$ for all $x\in\Omega$, the dominated convergence theorem gives
\bet
\tG(f_k) = \sum_{n=1}^k \tG(\chi_{K_n}f) = \sum_{n=1}^k \int (f|K_n) v_n d\nu_n = \int f_k v d\nu\goesto \int fv d\nu.
\eeqt
Thus,
\bet
\tG(f) = \int f v d\nu.
\eeqt
Since $\|\tG(f)\| \leq \int |f(x)|\|v(x)\| d\nu(x)\leq \|f\|_1M$ for all $f\in L^1(\Omega, \nu)$,
we get $\|\tG\|\leq M$, so $M=\|\tG\|$. Since $\nu$ is $\sigma$-additive, $v$ is $\nu$-essentially
unique by \cite[Corollary 5, p. 47]{Diestel}. The Lemma is proved. $\Box$

\

The next Proposition allows us to specify the nature of the function $v$ obtained in the previous Lemma, in the case
where $\Omega$ is a locally compact topological group possessing certain additional properties.
The next Lemma is essential to its proof.

\begin{lemma}\label{apulemma}
Let $\Omega$ be a locally compact second countable topological group with a left Haar measure $\nu$.
\begin{itemize}
\item[(a)] Let $h:\Omega\to \C$ be a $\nu$-measurable $\nu$-essentially bounded function such that for each
$y\in \Omega$, the function
$h(y\cdot)$ coincides with $h$ $\nu$-almost everywhere. Then there is a constant $c\in \C$, such that
$h(x) = c$ for $\nu$-almost all $x\in \Omega$.
\item[(b)] Let $X$ be a Banach space, and
$h:\Omega\to X$ a $\nu$-measurable $\nu$-essentially bounded function such that for each
$y\in\Omega$, the function $h(y\cdot)$ coincides with $h$ $\nu$-almost everywhere.
Then there is an $s\in X$, such that $h(x) =s$ for $\nu$-almost all $x\in \Omega$.
\end{itemize}
\end{lemma}
\noindent {\bf Proof. }
(a) Clearly the positive functions $h_i^{\pm}=\frac 12(|h_i|\pm h_i)$, $i=1,2$, for which $h=(h_1^+-h_1^-)+i(h_2^+-h_2^-)$,
share the property assumed to hold for $h$. Therefore, it suffices to prove the result in the case where
$h$ is positive.
Since $h$ is $\nu$-essentially bounded and $\nu$-measurable, the $\nu$-measurable function $fh$ is
$\nu$-integrable for all $f\in L^1(\Omega,\nu)$.
Let $\Cc(\Omega)$ denote the space of compactly supported continuous complex functions on $\Omega$.
We notice that the positive functional $I_h:\Cc(\Omega)\to \C$, defined by $I_h(f) = \int fhd\nu$, satisfies the relation
\bet
I_h(f) = \int f(x)h(x)d\nu(x) = \int f(yx)h(yx) d\nu(x) = \int f(yx)h(x) d\nu(x) = I_h(f(y\cdot))
\eeqt
for each $y\in \Omega$, and hence is a left Haar integral in the group $\Omega$. By the uniqueness theorem of
Haar integrals, there is a $c>0$, such that
$I_h(f) = c\int f d\nu$ for all $f\in \Cc(\Omega)$. Since $\Cc(\Omega)$ is dense in $L^1(\Omega,\nu)$, it follows that
$h(x) =c$ for almost all $x\in \Omega$.\\
(b) Fix some $A\in \hB(\Omega)$, such that $0 < \nu(A) <\infty$, and denote $s= \nu(A)^{-1}\int_A h d\nu\in X$.
Let $w^*\in X^*$. Since $h$ is $\nu$-measurable, so is the complex valued function $x\mapsto \bra w^*,h(x)\ket$, which thus
coincides almost everywhere with some Borel function $h_{w^*}$.
Since $(x,y)\mapsto xy$ is continuous, the function $(x,y)\mapsto h_{w^*}(xy)$ is
$\nu\times \nu$-measurable. By assumption, $h_{w^*}$ satisfies the conditions of (a), so there is a constant $c\in \C$,
and a $\nu$-null set $N$, such that $h_{w^*}(y)=c$ for all $y\in \Omega\setminus N$.
Let $x\in \Omega$. Since the left and right Haar measures have the same null sets, also $Nx^{-1}\cup x^{-1}N$ is
a $\nu$-null set. Thus, for each $x\in \Omega$, we have $h_{w^*}(yx)=c=h_{w^*}(xy)$ for almost all
$y\in \Omega$. Using this fact, the assumption and the Fubini-Tonelli theorem, we get for each $f\in L^1(\Omega,\nu)$,
\beat
\bra w^*, \nu(A)s\int f d\nu\ket &=& \int_A h_{w^*}(x) d\nu(x) \int f(y) d\nu(y)\\
&=& \int \left(\int \chi_A(x)h_{w^*}(x) f(y) d\nu(x)\right) d\nu(y)\\
&=& \int \left(\int \chi_A(x)h_{w^*}(yx) f(y) d\nu(x)\right) d\nu(y)\\
&=& \int \left(\int \chi_A(x)h_{w^*}(yx) f(y) d\nu(y)\right) d\nu(x)\\
&=& \int \left(\int \chi_A(x)h_{w^*}(xy) f(y) d\nu(y)\right) d\nu(x)\\
&=& \int \left(\int \chi_A(x)h_{w^*}(y)f(y) d\nu(y)\right) d\nu(x)= \bra w^*, \nu(A)\int f(y) h(y) d\nu(y)\ket.
\eeqat
The use of the Fubini-Tonelli theorem is justified because $\nu$ is $\sigma$-finite, 
$(x,y)\mapsto \chi_A(x)h_{w^*}(yx) f(y)$ is $\nu\times \nu$-measurable, and
\bet
\int \left(\int \|\chi_A(x)h_{w^*}(yx) f(y)\| d\nu(x)\right) d\nu(y) \leq \nu(A) \|w^*\|M\|f\|_1<\infty,
\eeqt
where $M>0$ is such that $\|h(x)\|\leq M$ for almost all $x\in \Omega$.
Since $w^*\in X^*$ was arbitrary, we get $\int_B h(y) d\nu(y) = \int_B s d\nu(y)$ for each $B\in \hB(\Omega)$ of finite measure.
Thus, from \cite[Corollary 5, p. 47]{Diestel} and the $\sigma$-finiteness of $\nu$
it follows that $h(x) = s$ for almost all $x\in \Omega$. $\Box$
  
\

\begin{proposition} \label{oplemmaB}
Assume that $\Omega$ is a locally compact second countable topological group with a left Haar measure $\nu$,
and $X$ a Banach space having the Radon-Nikod\'ym property.
In addition, assume that there is a homomorphism $\hol: \Omega\to \iso (X)$, such that
\begin{itemize} 
\item[(i)] $\sup_{x\in \Omega}\|\ho{x}\|<\infty$;
\item[(ii)] for all $w\in X$, the map $x\mapsto \ho{x^{-1}}(w)$ is $\nu$-measurable.
\end{itemize}
If $\tG:L^1(\Omega, \nu)\to X$ is a continuous linear map satisfying
$\ho{x} (\Gamma (f)) = \Gamma(f(x^{-1}\cdot))$ for all $f\in L^1(\Omega,\nu)$ and $x\in \Omega$,
then there is a unique vector $s\in X$, such that
\bet
\tG(f) =\int f(x) \ho{x}(s) d\nu(x)
\eeqt
for all $f\in L^1(\Omega, \nu)$. If each $\ho{x}$ is an isometry, then $\|s\| = \|\tG\|$.
\end{proposition}
\noindent {\bf Proof. }
Let $v:\Omega\to X$ be the function obtained in Lemma \ref{oplemmaA}. We have to prove that for some unique $s\in X$, it satisfies
$v(x) = \ho{x} (s)$ for almost all $x\in \Omega$.
To that end, let $B\in \hB(\Omega)$ be such that $\nu(B)<\infty$, and $y\in \Omega$. Then, by the continuity of the linear map
$\ho{y}$ we get, by using the left invariance of $\nu$, 
\beat
\int_B \ho{y}(v(x)) d\nu(x) &=& \ho{y}(\tG(\chi_B))= \tG(\chi_B(y^{-1}\cdot))
= \int \chi_B(y^{-1}x) v(x) d\nu(x) \\
&=&\int_B v(yx) d \nu(x).
\eeqat
Since the measure $\nu$ is $\sigma$-finite, it follows from \cite[Corollary 5, p. 47]{Diestel} that
for each $y\in \Omega$,
\be\label{tcov}
\ho{y}(v(x))= v(yx)\text{ for almost all } x\in \Omega.
\eeq

Now define the map $v_0: \Omega\to X$ by $v_0(x) = \ho{x^{-1}} (v(x))$.
Then $v_0$ is $\nu$-measurable. Indeed, let $B\in\hB(\Omega)$ be
such that $\nu(B) <\infty$. Since $v$ is $\nu$-measurable, there is a sequence
$(v_n)$ of $\nu$-simple functions vanishing outside $B$ and converging $\nu$-a.e. to $\chi_Bv$.
For each $w\in X$, the map $x\mapsto \ho{x^{-1}}(w)$ is $\nu$-measurable by
assumption (ii), so that also the functions $x\mapsto \ho{x^{-1}}(v_n(x))$, are $\nu$-measurable.
Now $\ho{x^{-1}}(v_n(x))\goesto \chi_B(x) \ho{x^{-1}}(v(x))=\chi_Bv_0(x)$ for $\nu$-almost all $x$, because
$\ho{x^{-1}}$ is continuous, so the limit $\chi_Bv_0$ is $\nu$-measurable \cite[p. 150]{Dunford}. Thus $v_0$ is
$\nu$-measurable.

Let $f\in L^1(\Omega, \nu)$. Since $v_0$ is $\nu$-measurable, so is $fv_0$ \cite[p. 106]{Dunford}. In addition,
since $\sup_{x\in \Omega}\|v(x)\| = \|\Gamma\|$, we get $\|f(x)v_0(x)\|\leq M|f(x)|\|v(x)\|\leq |f(x)| M\|\tG\|$ for all
$x$, where $M=\sup_{x\in \Omega}\|\ho{x}\|<\infty$, so $fv_0$ is $\nu$-integrable.
In particular, $v_0$ is integrable over any set $B\in \hB(\Omega)$ of finite measure. Also, $\|v_0(x)\|\leq M\|\tG\|$ for all
$x$, so $v_0$ is $\nu$-essentially bounded.
Since $\hol$ is a homomorphism, $v(x) =\ho{x}(v_0(x))$ for all $x$. Let $y\in \Omega$. The result (\ref{tcov}) gives
$\ho{y}(\ho{x}(v_0(x))) = \ho{yx}(v_0(yx))$ for almost all $x$, so that 
\be\label{cov0}
\text{for each } y\in \Omega, \ v_0(x) = v_0(yx) \text{ for almost all } x\in \Omega.
\eeq
By Lemma \ref{apulemma} there is an $s\in X$, such that $v_0(x) = s$ for $\nu$-almost all $x$.
Thus
\bet
\tG(f) = \int f(x) v(x) d\nu(x)=\int f(x) \ho{x} (v_0(x)) d\nu(x)=\int f(x) \ho{x} (s) d\nu(x)
\eeqt
for all $f\in L^1(\Omega, \nu)$. The vector $s$ in the above representation is uniquely determined, because if
$s'\in X$ had the same properties, then by the uniqueness of the map $x\mapsto v(x)$ in the representation
of Lemma \ref{oplemmaA}, $\ho{x}(s) = v(x) = \ho{x}(s')$ for almost all $x\in \Omega$, so that $s=s'$.

If $\ho{x}$ is an isometry for each $x\in \Omega$, we have in addition,
\bet
\|\tG\| = \sup_{x\in \Omega} \|v(x)\|=\sup_{x\in \Omega} \|\ho{x} (s)\| = \|s\|
\eeqt
The proof is complete. $\Box$

\section{Positive normal covariant maps}\label{sec4}

Now we return to the concept of $(G, \holg, d)$ introduced earlier.
Theorem \ref{Theorem2} below characterizes all positive normal $\holg$-covariant maps $\Gamma:L^\infty(G, \lambda)\to L(\hil)$.
The proof is based on the fact that $\hT(\hil)$, being a separable dual space, has the Radon-Nikod\'ym
property by \cite[p. 79]{Diestel}. Therefore, the following Lemma is needed. We give the proof for completeness. (The result is
given without proof e.g. in \cite[Exercise 5.7, p. 131]{Stratila}.)

\begin{lemma}\label{separabilitylemma}
The space $\hT(\hil)$ is separable (with respect to the trace norm).
\end{lemma}
\noindent {\bf Proof. } If $\vp,\psi\in\hil$ are such that $\|\vp\|=1$, and $\|\vp-\psi\|\leq 1$, then
\be\label{normineq}
\||\psi\ket\bra\psi|-|\vp\ket\bra\vp|\|_{\rm tr}\leq 3\|\psi-\vp\|.
\eeq
Indeed, since the map $\hT(\hil)\ni T\mapsto \tr[T\cdot]\in \h C(\hil)^*$, where $\h C(\hil)$ denotes the set of compact
operators, is an isometry, we have $\|T\|_{\rm tr}=\sup \{ |\tr[TA]|\mid A\in \h C(\hil), \|A\| \leq 1\}$ for each $T\in\hT(\hil)$.
Let $\vp,\psi\in\hil$ be such that $\|\vp\|=1$, and $\|\vp-\psi\|\leq 1$. If $A\in \h C(\hil)$, $\|A\| \leq 1$, we have
\beat
|\tr [(|\psi\ket\bra\psi|-|\vp\ket\bra\vp|)A]| &=& |\bra \psi |A\psi\ket-\bra \vp |A\vp\ket|
\leq |\bra \psi|A\psi\ket-\bra\psi|A\vp\ket|+|\bra\psi |A\vp\ket-\bra\vp|A\vp\ket| \\
&\leq& \|\psi\|\|\psi-\vp\|+\|\psi-\vp\|\|\vp\|\\
&\leq& (\|\psi-\vp\|+\|\vp\|)\|\psi-\vp\|+\|\psi-\vp\|\|\vp\| \leq 3\|\psi-\vp\|.
\eeqat
Thus (\ref{normineq}) holds.

Let $M$ be a countable dense set in the separable space $\hil$.
Define $\hF$ to be the set of operators of the form $\sum_{\psi\in F} \lambda_{\psi} |\psi\ket\bra\psi|$, where
$F$ is a finite subset of $M$ and each $\lambda_{\psi}$ is a positive rational number (the vectors $\psi$ need not be
of unit length). Since $M$ and $\Q$ are
countable sets, $\hF$ is countable. Clearly $\hF$ is a subset of the set $\hT(\hil)^+$ of positive trace-class operators.
We proceed to show that $\hF$ is $\|\cdot\|_{\rm tr}$-dense in $\hT(\hil)^+$.

Let $S\in\hT(\hil)^+$ and $\epsilon>0$. Using the decomposition $S=\sum_n t_n |\vp_n\ket\bra\vp_n|$,
which converges in the trace norm, with $t_n\geq 0$ and the $\vp_n$ orthonormal unit vectors, we find that there is a
$k\in\N$, such that
\be\label{eq1}
\left\|S-\sum_{n=1}^k t_n |\vp_n\ket\bra\vp_n|\right\|_{\rm tr}< \frac{\epsilon}{3}.
\eeq
Now we choose positive rational numbers $\lambda_n$, $n=1,\ldots, k$, such that
$|t_n-\lambda_n|< \frac {\epsilon}{3k}$ for all $n=1,\ldots, k$.
Then
\be\label{eq2}
\left\| \sum_{n=1}^k t_n |\vp_n\ket\bra\vp_n|-\sum_{n=1}^k \lambda_n |\vp_n\ket\bra\vp_n|\right\|_{\rm tr}< \frac {\epsilon}{3}.
\eeq
Since $M$ is dense, we can pick vectors $\psi_n\in M$, $n=1,\ldots,k$, such that
$\|\psi_n -\vp_n\|< \frac {\epsilon}{9\sum_{n=1}^k\lambda_n}$ for all $n=1,\ldots, k$. It can be assumed that
$\epsilon <1$, so that we can use the result (\ref{normineq}) to get
\be\label{eq3}
\left\|\sum_{n=1}^k \lambda_n |\vp_n\ket\bra\vp_n|- \tilde{S}\right\|_{\rm tr} <\frac {\epsilon}{3},
\eeq
where $\tilde{S}=\sum_{n=1}^k \lambda_n |\psi_n\ket\bra\psi_n|\in \hF$.
The inequalities (\ref{eq1})-(\ref{eq3}) now imply $\|S-\tilde{S}\| <\epsilon$. Thus $\hF$ is $\|\cdot\|_{\rm tr}$-dense in
$\hT(\hil)^+$.

Since $\hT(\hil) = \hT(\hil)^+-\hT(\hil)^++i(\hT(\hil)^+-\hT(\hil)^-)$, the set $\hF-\hF+i(\hF-\hF)$ is a countable dense subset
of $\hT(\hil)$. $\Box$

\begin{theorem}\label{Theorem2}
Let $\Gamma:L^\infty(G, \lambda)\to L(\hil)$ be a normal positive $\holg$-covariant linear map satisfying $\Gamma (g\mapsto 1) = I$. 
Then $\Gamma$ is of the form of Theorem \ref{Theorem1} for a unique positive operator $T\in \hT(\hil)$ of trace one.
\end{theorem}
\noindent {\bf Proof. } Since $\Gamma: L^1(G, \lambda)^*\to \hT(\hil)^*$ is a weak-* continuous linear map, there is
a linear map $\Gamma_*:\hT(\hil)\to L^1(G, \lambda)$, such that $(\Gamma_*)^*= \Gamma$.
The map $\Gamma_*$ is also positive, since
$\int (\Gamma_*(S))(g)f(g)d\lambda(g) = \tr [\Gamma(f)S] \geq 0$ for all positive $S\in\hT(\hil)$ and $f\in L^\infty(G, \lambda)$,
$f\geq 0$. Let $S\in\hT(\hil)$ be positive and $f\in L^\infty(G, \lambda)\cap L^1(G, \lambda)$ a positive function. Then
$\Gamma(f)$ is a positive operator and $\Gamma_*(S)$ a positive function. By covariance, we have
\bet
\tr [\Gamma(f) \hog{g}(S)] = \tr [\hog{g}^* (\Gamma(f))S] = \tr [S\Gamma(f(g\cdot))]
= \int (\Gamma_*(S))(g')f(gg')d\lambda(g'),
\eeqt
from which it follows by the Fubini-Tonelli theorem and the right invariance of $\lambda$ that 
\bet
\int \tr [\Gamma(f)\hog{g}(S)] d\lambda(g) = \int (\Gamma_*(S))(g') \left(\int f(gg') d\lambda(g)\right)d\lambda(g')
= \|\Gamma_*(S)\|_1 \|f\|_1<\infty.
\eeqt
Lemma \ref{trlemma} now implies that $\Gamma(f)\in \hT(\hil)$, and since
$\|\Gamma_*(S)\|_1 = \int \Gamma_*(S)(g)d\lambda(g) = \tr [S \Gamma(g\mapsto 1)] = \tr [S]$, we find (by using Lemma \ref{trlemma} again) that
for positive $f\in L^\infty(G, \lambda)\cap L^1(G, \lambda)$, $\|\Gamma(f)\|_{\rm tr} = d^{-1}\|f\|_1$. If
$f\in L^1(G, \lambda)\cap L^\infty(G, \lambda)$ is
arbitrary, we can write $f= (f_1^+-f_1^-)+i(f_2^+-f_2^-)$, where the $f_i^{\pm}$ are positive, and
$f_1^++f_1^-+f_2^++f_2^- = |f_1|+|f_2|\leq 2|f|$. It then follows by the linearity of $\Gamma$ that
$\|\Gamma(f)\|_{\rm tr} \leq 2d^{-1}\|f\|_1$, implying that the restriction
$\Gamma| L^1(G, \lambda)\cap L^\infty(G, \lambda):L^1(G, \lambda)\cap L^\infty(G, \lambda)\to \hT(\hil)$ is continuous
with respect to the norms $\|\cdot\|_1$ and $\|\cdot\|_{\rm tr}$.
Since the set $L^1(G, \lambda)\cap L^\infty(G, \lambda)$ contains all integrable simple functions, it is dense in
$L^1(G, \lambda)$. Therefore (since $\hT(\hil)$ is complete), $\Gamma|L^1(G, \lambda)\cap L^\infty(G, \lambda)$ can be extended to a
continuous linear map $\wG:L^1(G, \lambda)\to \hT(\hil)$.

The map $\wG$ is positive. In fact, if $f\in L^1(G, \lambda)$
is positive, there is an increasing sequence $(f_n)$ of integrable positive simple functions converging pointwise to $f$.
By the monotone convergence theorem, $f_n\goesto f$ in the $\|\cdot\|_1$-norm, so that the trace-class operator
$\wG(f)$, being the trace-norm (and hence weak) limit of the sequence $\Gamma(f_n)$ of positive trace-class operators,
must be positive.

Now we show that the conditions of Proposition \ref{oplemmaB} are satisfied by the measure space $(G, \hB(G), \lambda)$,
the Banach space $\hT(\hil)$, the homomorphism $\holg$, and the linear map $\wG$.

Since $\hT(\hil)\cong\h C(\hil)^*$ is separable by Lemma \ref{separabilitylemma},
it has the Radon-Nikod\'ym property \cite[p. 79]{Diestel}.
Since each $\hog{g}$ is an isometry, the condition (i) is holds. 
Let $S\in \hT(\hil)$ and $A\in L(\hil)$. Since $\lambda$ is a Borel measure, the map $g\mapsto \tr[A\hog{g^{-1}}(S)]$,
being continuous, is also $\lambda$-measurable.
Thus $G\ni g\mapsto w^*(\hog{g^{-1}}(S))\in \C$ is $\lambda$-measurable for each $w^*\in \hT(\hil)^* \cong L(\hil)$. Since $\hT(\hil)$ is
separable, this implies by \cite[p. 149]{Dunford} that the map
$g\mapsto \hog{g^{-1}}(S)$ is measurable, so that the condition (ii) of Proposition \ref{oplemmaB} is satisfied.
To verify condition (iii), let $f\in L^1(G, \lambda)$, $g\in G$. Choose a sequence $(f_n)$ of
integrable simple functions converging to $f$ in the $\|\cdot\|_1$-norm. 
Thus, by the continuity of the mappings involved, the covariance of $\Gamma$, and the fact that the
map $\hog{g^{-1}}^* = (\hog{g}^{-1})^*$ coincides with $\hog{g}$ on $\hT(\hil)$, we get
\bet
\hog{g} (\wG(f)) = \lim_n \hog{g^{-1}}^* (\Gamma(f_n)) = \lim_n \Gamma(f_n(g^{-1}\cdot)) = \wG(f(g^{-1}\cdot)),
\eeqt
where the limits are in the trace norm and the $\|\cdot\|_1$-norm. This proves that (iii) holds.

Thus, we can apply Proposition \ref{oplemmaB} to the map $\wG$. There is a unique $T'\in \hT(\hil)$, such that
\bet
\wG(f) = \int f(g) \hog{g} (T') d\lambda(g)
\eeqt
for all $f\in L^1(G, \lambda)$.
Since $L^\infty(G,\lambda)\cap L^1(G,\lambda)$ is weak-* dense in $L^\infty(G,\lambda)$ and $\Gamma$
is normal, we also have
\bet
\Gamma(f) = \int f(g) \hog{g} (T') d\lambda(g)
\eeqt
in the ultraweak sense for all $f\in L^\infty(G, \lambda)$.

It remains to prove that $T'$ is positive and of trace $d^{-1}$.

Let $S\in\hT(\hil)$ be positive. Since $\Gamma(\chi_{B})$ is a positive operator, we have
\bet
0\leq \tr[S\Gamma(\chi_{B})] = \int_B \tr [S\hog{g} (T')] d\lambda(g)
\eeqt
for all $B\in \hB(G)$, from which it follows by the continuity of $g\mapsto \tr [S\hog{g} (T')]$
that $\tr [S\hog{g} (T')]\geq 0$ for all $g\in G$. Thus $T'$ must be positive.

In addition, by the condition $\Gamma(g\mapsto 1) = I$, and
Lemma \ref{trlemma},
\bet
\tr [S]d^{-1}= d^{-1}\tr [S \Gamma(\chi_G)] = d^{-1}\int \tr [S\hog{g} (T')] d\lambda(g) = \tr [S] \tr [T']
\eeqt
for any positive $S\in \hT(\hil)$. Thus $\tr [T'] = d^{-1}$, so that by defining $T=T'd$, we get the required form for $\Gamma$.
 $\Box$

\

\section{Covariant observables}\label{sec5}

An observable (i.e. a positive normalized operator measure) $E:\hB(G)\to L(\hil)$ is said to be \emph{$\holg$-covariant} if
$\hog{g}^* (E(B)) = E(g^{-1}B)$ for all $g\in G$ and $B\in \hB(G)$.
The following Lemma shows that Theorem \ref{Theorem2} can be used to characterize the covariant observables.
The result (b) of the Lemma is obtained in \cite{norm1} for the more general case where the group need not be unimodular,
and the condition (\ref{prop1}) is not assumed. In the context of this paper, the
proof following \cite{Holevo} is more simple, as it can be formulated so that it uses Lemma \ref{trlemma}. The proof is therefore
given here. 
\begin{lemma}\label{contlemma} Let $E:\hB(G)\to L(\hil)$ be an observable.
\begin{itemize}
\item[(a)] Assume that for each trace class operator $S$, the measure $B\mapsto \tr[S E(B)]$ is continuous with respect to
the measure $\lambda$. Then for each $f\in L^\infty(G, \lambda)$, the operator integral $\int f dE$ exists in $L(\hil)$ in
the ultraweak sense,
and the linear map $f\mapsto \int f dE$ is normal, positive, and satisfies $\int 1dE(g) = I$.
If $E$ is $\holg$-covariant, so is the map $f\mapsto \int f dE$.
\item[(b)] If $E$ is $\holg$-covariant, the measure $B\mapsto \tr[S E(B)]$ is continuous with respect to
the measure $\lambda$ for each trace class operator $S$.
\end{itemize}
\end{lemma}
\noindent {\bf Proof. }
(a) Let $S\in \hT(\hil)$. Then $S=\sum_n t_n |\psi_n\ket\bra\vp_n|$, where $(\vp_n)$ and $(\psi_n)$
are orthonormal sequences, $t_n\geq 0$, and $\sum t_n < \infty$. The series converges in the trace norm.
The map $\mu$, defined by $B\mapsto \mu(B) = \tr[SE(B)]$ is a
complex valued finite measure, and (by the $\|\cdot\|_{\rm tr}$-continuity of the trace functional)
it is a pointwise limit of the measures $\sum_{n=1}^k\mu_n$, where
$\mu_n(B) = t_n \tr [|\psi_n\ket\bra \vp_n|E(B)]$ for each $B\in \hB(G)$. Since the total variation norm of
$\mu_n$ satisfies $\| \mu_n\| \leq 4 \sup_{B\in \hB(G)} |\mu_n(B)|\leq 4t_n$,
the series $\mu=\sum_n \mu_n$ converges absolutely in the total variation norm.

Let $f\in L^\infty(G, \lambda)$. Since $\mu$ and each $\mu_n$ are $\lambda$-continuous, $|f(g)|\leq \|f\|_\infty$ also $\mu_n$-,
and $\mu$-almost everywhere. Thus, $\int |f| d|\mu_n|\leq \|f\|_\infty \|\mu_n\| \leq 4\|f\|_\infty t_n$ so that
$\sum_n \int |f| d|\mu_n| \leq 4 \|f\|_\infty \sum_n t_n = 4\|f\|_\infty \|S\|_{\rm tr}< \infty$.
It now follows e.g. from \cite[Lemma 1]{Momentarticle} that $f$ is $\mu$-integrable, and
\bet
\int f d (\tr[SE(\cdot)]) = \int f d\mu = \sum_n \int f d\mu_n = \sum_n t_n \int f d(\tr[|\psi_n\ket\bra \vp_n|E(\cdot)]).
\eeqt
Since $\mu$ is $\lambda$-continuous, the integral does not depend on the representative of $f\in L^\infty(G,\lambda)$.
In addition,
\be\label{cont}
\left|\int f d (\tr[SE(\cdot)])\right|\leq \sum_n \int |f| d|\mu_n| = 4\|f\|_\infty \|S\|_{\rm tr},
\eeq
so that the functional $S\mapsto \int f d (\tr[SE(\cdot)])$ is $\|\cdot\|_{\rm tr}$-continuous. Thus
the integral $\int f dE$ exists in the ultraweak sense as an operator in $L(\hil)$, i.e., for each $S\in\hT(\hil)$,
\be\label{ultra}
\tr[S (\int f dE)] = \int f d(\tr[SE(\cdot)]).
\eeq
Since $B\mapsto \tr[SE(B)]$ is $\lambda$-continuous, it has a density $g_S\in L^1(G, \lambda)$. Since
$L^\infty(G, \lambda)\ni f\mapsto \int f dE\in L(\hil)$ is the dual map of $\hT(\hil)\ni S\mapsto g_S\in L^1(G, \lambda)$, it is
normal.

Let $f\in L^\infty(G, \lambda)$ be positive and $S\in\hT(\hil)$ a positive operator. Since the measure $\tr[S E(\cdot)]$ is positive,
so is $\tr[S (\int f dE)] = \int f d(\tr[S E(\cdot)])$. It follows that $\int f dE$ is positive.
Thus the map $f\mapsto \int f dE$ is positive. Since $E$ is normalized, $\int 1 dE(g) = E(G) = I$.

Assume now that $E$ is $\holg$-covariant.
Let $g\in G$, $B\in \hB(G)$, and $S\in \hT(\hil)$. Since the measure $\tr[SE(\cdot)]$
has the density $g_S\in L^1(G, \lambda)$, the measure $\tr[SE(g^{-1}\cdot)]$ has the density $g_S(g^{-1}\cdot)$.
Using the left invariance of $\lambda$ and the covariance of $E$, we get
\beat
\tr[S\hog{g}^*(\int f dE)] &=& \tr[\hog{g} (S)(\int f dE)] = \int f d(\tr[\hog{g}(S)E(\cdot)])
=  \int f d(\tr[SE(g^{-1}\cdot)])\\
&=& \int f(g') g_S(g^{-1}g') d\lambda(g') = \int f(gg') g_S(g') d\lambda(g') = \int f(g\cdot) d(\tr[SE(\cdot)]) \\
&=& \tr[S (\int f(g\cdot) dE)],
\eeqat
which proves that the map $f\mapsto \int f dE$ is $\holg$-covariant.

(b) Let $S\in\hT(\hil)$ be positive and of trace one, and $\mu$ the probability measure
$B\mapsto \tr[S E(B)]$. Now for each $B\in \hB(G)$, covariance implies
\bet
\tr[\hog{g}^* (E(B))S] = \tr[SE(g^{-1}B)] = \int \chi_{g^{-1}B}d\mu = \int \chi_{B}(gg')d\mu(g').
\eeqt  
Thus, by Lemma \ref{trlemma}, the Fubini-Tonelli theorem, and the right invariance of $\lambda$, we get
\beat
\tr [E(B)] &=& d^{-1}\int \tr[E(B) \hog{g} (S)] d\lambda(g) = d^{-1}\int \left(\int\chi_B(gg') d\lambda(g)\right) d\mu(g')\\
&=& d^{-1}\lambda(B) \int d\mu = d^{-1}\lambda(B).
\eeqat
Now let $S\in \hT(\hil)$ be arbitrary. Then, if $B\in \hB(G)$ is such that $\lambda(B)<\infty$, we have
$|\tr[S E(B)]| \leq \|S\| \|E(B)\|_{\rm tr} = d^{-1}\|S\| \lambda(B)$. This implies that the measure $B\mapsto \tr[S E(B)]$ is
$\lambda$-continuous. $\Box$

\begin{theorem}\label{observables}
Let $E:\hB(G)\to L(\hil)$ be a positive normalized $\beta$-covariant operator measure. Then
\bet
E(B) = d^{-1}\int_B \hog{g} (T) d\lambda(g)
\eeqt
in the ultraweak sense, for some unique positive operator $T\in \hT(\hil)$ of trace one.
\end{theorem}
\noindent {\bf Proof. }
By the previous Lemma, the linear map $L^\infty(G, \lambda)\ni f\mapsto \int f dE\in L(\hil)$ satisfies the
conditions of Theorem \ref{Theorem2} and hence is of the form
\bet
\int f dE = d^{-1}\int f(g) \hog{g} (T) d\lambda(g)
\eeqt
for some unique positive operator $T$ of trace one. In particular,
\be\label{Erep}
E(B) = \int \chi_B dE = d^{-1}\int_B \hog{g} (T) d\lambda(g)
\eeq
for each $B\in \hB(G)$. The operator $T$ in the representation (\ref{Erep}) of $E$ is also uniquely determined. In fact, if $S\in\hT(\hil)$ is such that
$E(B) = d^{-1}\int_B\hog{g}(S) d\lambda(g)$ for each $B\in \hB(G)$, then by the uniqueness of $T$ in the representation
of the linear map $f\mapsto \int f dE$,
we get $\int \chi_B(g) \hog{g}(S)d\lambda(g) = \int \chi_B(g) \hog{g}(T)d\lambda(g)$ for all $B\in \hB(G)$, so
$\hog{g}(S)=\hog{g}(T)$ for almost all $g$, showing that $S=T$. $\Box$

\

\noindent {\bf Remark. } Consider the concrete case $(\R^{2n}, \holc, (2\pi)^n)$.
For a linear map $\Gamma:L^\infty(\R^{2n}, \mu_L)\to L(L^2(\R^n))$, covariance
means that $\hoc{x}(\Gamma(f)) = f(\cdot -x)$ for all $x\in \R^{2n}$ and $f\in L^\infty(\R^{2n}, \mu_L)$, whereas a covariant observable
$E:\hB(\R^{2n}) \to L(L^2(\R^n))$ is such that $\hoc{x}(E(B)) = E(x+B)$ for each $x\in \R^{2n}$ and $B\in \hB(\R^{2n})$. Thus Theorem \ref{Theorem2}
gives, in particular,  a characterization of positive covariant linear maps
$\Gamma:L^\infty(\R^{2n}, \mu_L)\to L(L^2(\R^n))$, and Theorem
\ref{observables} a characterization of the covariant phase space observables.

\section{ A note on quantization maps on the set of unbounded functions} \label{sec6}
Since many of the important dynamical variables in classical mechanics are unbounded functions,
it is rather restrictive to consider only the quantization maps $\Gamma:L^\infty(G,\lambda)\to L(\hil)$.

Let $\hF(G)$ denote the set of all complex Borel functions on $G$, and $\uop$ the set of all (not necessarily bounded)
linear operators in $\hil$. We call a map $\Gamma:\hF(G)\to\uop$ linear if
$\alpha\Gamma(f)+\beta\Gamma(h)\subset \Gamma(\alpha f+\beta h)$ for all $\alpha,\beta\in\C$ and $f,h\in \hF(G)$.
For each $f\in \hF(G)$, we let $D(\Gamma(f))$ denote the domain of $\Gamma(f)$.
 
Let  $E:\hB(G)\to L(\hil)$ be a positive operator measure.
For $f\in\hF(G)$ let $\intd{f}{E}$ be the set of those vectors $\vp\in\hil$ for which $f$ is $E_{\psi, \vp}$-integrable
for all $\psi\in\hil$.
The operator integral $\intL{f}{E}=\int f dE$ is defined to be the unique
(possibly unbounded) linear operator on the domain $\intd{f}{E}$, for which
$\bra\psi |\intL{f}{E}\vp\ket = \int f dE_{\psi, \vp}$ for all $\vp\in \intd{f}{E}$ and $\psi\in\hil$ (cf. \cite{Lahti}).
If $f$ is real valued, then $\intL{f}{E}$ is a symmetric operator.

Consider the map $\Gamma_E:\hF(G)\to \uop$, defined by $\Gamma_{E}(f)= L(f, E)$. If $f,h\in \hF(G)$,
$\alpha, \beta\in \C$, then (since $|f+h|\leq |f|+|h|$) $\alpha\Gamma(f)+\beta\Gamma(h)\subset \Gamma(\alpha f+\beta h)$,
so $\Gamma_E$ is linear. It follows from the dominated convergence theorem that it is quasicontinuous in the sense
of the following definition (already given in the Introduction).

\

{\bf Definition.} A linear map $\Gamma:\hF(G)\to \uop$ is \emph{quasicontinuous}, if for each increasing sequence
$(f_n)$ of positive Borel functions converging pointwise to an $f\in \hF(G)$ the numerical sequence
$(\bra \psi |\Gamma(f_n)\vp\ket)$ converges to $\bra \psi |\Gamma(f)\vp\ket$ for all $\psi\in \hil$ and 
$\vp\in D(\Gamma(f))\cap\bigcap_{n\in N} D(\Gamma(f_n))$.

\

In the Introduction we mentioned that in order to be represented as an operator integral, a quantization map $\Gamma$
must be at least positive, linear and quasicontinuous, and map bounded functions to $L(\hil)$, for then
the map $E^{\Gamma}:\hB(G)\to L(\hil)$, given by $B\mapsto \Gamma(\chi_B)$ is a positive operator measure,
and $\Gamma(f) = L(f, E^{\Gamma})$ for each bounded function $f\in \hF(G)$.
In order to claim that $\Gamma = L(\cdot, E^{\Gamma})$, something must be assumed on the domains of the operators $\Gamma(f)$.
The following simple result follows readily from the definition of the operator integral:

\begin{proposition}
Let $\Gamma:\hF(G)\to \uop$ be a linear map satisfying the following conditions:
\begin{itemize}
\item[(i)] $\Gamma$ is positive and quasicontinuous;
\item[(ii)] $\Gamma$ maps bounded functions to $L(\hil)$;
\item[(iii)] for $f\in\hF(G)$, the domain of $\Gamma(f)$ consists of those vectors $\vp\in \hil$
for which $f$ is $E^{\Gamma}_{\psi,\vp}$-integrable for all $\psi\in\hil$.
\end{itemize}
Then $\Gamma(f)= L(f, E^{\Gamma})$ for all $f\in \hF(G)$.
\end{proposition}
{\bf Proof. } As (iii) asserts that the domains of the operators $\Gamma(f)$ and $L(f, E^{\Gamma})$ are the same,
we are left to show that $\Gamma(f)\vp=L(f,E^{\Gamma})\vp$ for all $\vp$ in the common domain $\hD$.
Let $f\in \hF(G)$, $\vp\in \hD$ and $\psi\in\hil$. Assume first that $f$ is positive.
Pick an increasing sequence $(f_n)$ of $\hB(G)$-simple functions converging pointwise to $f$. By (ii),
$D(\Gamma(f_n)) =\hil$, so quasicontinuity implies that the sequence $(z^{\psi}_n)$, where
$z_n^{\psi}=\bra \psi|\Gamma(f_n)\vp\ket$, converges to $\bra \psi |\Gamma(f)\vp\ket$
for all $\psi\in \hil$. Since each $f_n$ is bounded, $\Gamma(f_n) = L(f_n, E^{\Gamma})$ for all
$n\in\N$, so $z_n^{\psi} = \int f_n d E^{\Gamma}_{\psi, \vp}$. But now (iii) and the dominated
convergence theorem imply that $z_n^{\psi}$ converges to
$\int f dE^{\Gamma}_{\psi, \vp} = \bra \psi | L(f, E^{\Gamma})\vp\ket$, so
$\bra \psi |\Gamma(f)\vp\ket=\bra \psi | L(f, E^{\Gamma})\vp\ket$. Since $\psi\in\hil$ was arbitrary,
this gives $\Gamma(f)\vp=L(f, E^{\Gamma})\vp$. For a general $f\in \hF(G)$, we
write $f=f_1^+-f_1^-+i(f_2^+-f_2^-)$, where $f_j^{\pm}$ are the positive and negative parts of $f_j$.
Let $\vp\in\hD$ and $\psi\in\hil$. Since $0\leq f_j^{\pm}\leq |f|$, we have that also
$f_j^{\pm}$ is $E^{\Gamma}_{\psi,\vp}$-integrable for all $\psi\in\hil$, i.e.
$\vp\in \intd{f_j^{\pm}}{E^{\Gamma}}=D(\Gamma(f_j^{\pm}))$.
Thus, $\Gamma(f_j^{\pm})\vp = L(f_j^{\pm}, E^{\Gamma})\vp$. By linearity, we get
$\Gamma(f)\vp = L(f, E^{\Gamma})\vp$, completing the proof. $\Box$

\end{document}